\documentclass[conference]{IEEEtran}
\usepackage{amsmath}
\usepackage{graphicx}
\usepackage{amsmath,amssymb}
\usepackage{bm}
\usepackage{url}
\usepackage{ascmac}

\usepackage{comment}

\title{Proximal Decoding \\ for LDPC-coded Massive MIMO Channels}

\author{%
  \IEEEauthorblockN{
  		Tadashi Wadayama
  		and Satoshi Takabe}
  \IEEEauthorblockA{\IEEEauthorrefmark{1}%
		Nagoya Institute of Technology,
		Gokiso, Nagoya, Aichi 466-8555, Japan,\\
 		\{wadayama, s\_takabe\}@nitech.ac.jp}
}

\begin{document}
%
\maketitle

\begin{abstract}
We propose a novel optimization-based decoding
algorithm for LDPC-coded massive MIMO channels.
The proposed decoding algorithm is based on 
a proximal gradient method for solving an approximate maximum a posteriori (MAP) decoding problem.
The key idea is the use of a code-constraint polynomial
penalizing a vector far from a codeword
as a regularizer in the approximate MAP objective function.
The code proximal operator is naturally 
derived from code-constraint polynomials.
The proposed algorithm, called proximal decoding,  can be described by a simple recursion 
consisting of the gradient descent step for a negative log-likelihood function and 
the code proximal operation.
Several numerical experiments show that the proposed algorithm outperforms 
known massive MIMO detection algorithms, such as an MMSE detector with 
belief propagation decoding.
\end{abstract}

\section{Introduction}

Low-density parity-check (LDPC) codes have been widely used  
in practical communications and storage systems,
such as mobile wireless communications, digital satellite broadcasting, optical communications, 
hard disks, and flash memories.
For decoding LDPC codes, belief propagation (BP) decoding is the de facto 
standard, but, in some cases,  {\em optimization-based decoding} algorithms have  
attracted the interest of researchers \cite{Feldman} \cite{Draper}.
A gradient descent formulation of a non-convex objective function 
including a penalty function for codewords leads to 
the concept of a gradient descent bit-flipping (GDBF) algorithm \cite{wadayama10},
which is suitable for hardware implementation requiring high-speed processing.
A number of variants of the GDBF algorithm have been proposed, and some 
of these variants, especially the noisy GDBF algorithm \cite{noisyGDBF}, provide excellent trade-offs between 
decoding performance and circuit complexity.
Another advantage of optimization-based decoding algorithms is 
that these algorithms can be applied to a more general channel model, including 
channels with memory \cite{interiorpoint}. 
Note that BP decoding is derived based on the memoryless property of the target channel.
This means that applying BP to channels with memory is not a trivial problem.
It may be possible to formulate maximum a posteriori (MAP) decoding for channels with memory as a 
non-convex optimization problem. 

In the present paper, we investigate a new direction for optimization-based decoding 
based on a {\em proximal gradient method} \cite{proximal}. The proximal gradient method is a well-known iterative minimization algorithm for convex optimization problems. For example, the 
iterative soft-thresholding algorithm (ISTA) \cite{ista} is an efficient sparse 
signal recovery algorithm, which is an instance of the proximal gradient method.
The proposed algorithm, referred to as {\em proximal decoding}, is conceptually 
very similar to the ISTA. 
The key idea is the use of the code-constraint polynomials
penalizing a vector far from a codeword
as a regularizer in the approximate MAP decoding.
The code proximal operator is naturally 
derived from code-constraint polynomials, which is the most important 
ingredient of the proposed method. The main contributions of the present paper are
1) a new formulation of an optimization-based decoding, i.e., proximal decoding,   
and 2) demonstrating that proximal decoding is competitive with known decoding algorithms in 
time complexity and in bit error rate (BER) performance over LDPC-coded Massive MIMO channels.

In the present paper, we focus on 
massive MIMO channel as a target channel because decoding and detection problems
for LDPC-coded massive MIMO channels are nontrivial problems and are also practically 
important problems \cite{massive} for 
wireless cellular networks referred to as fifth-generation (5G) systems, 
as well as for future systems such as beyond 5G/6G systems.
The authors recently proposed a detection algorithm 
for overloaded massive MIMO channels \cite{tpg}, and the architecture of the detection algorithm proposed in the previous study \cite{tpg} is another trigger 
for the development of proximal decoding. 

\section{Code-constraint polynomial}

\subsection{Notation}

Let $n$ be a positive integer representing code length.
A binary matrix $\bm H \in \mathbb{F}_2^{m \times n}$
is a parity check matrix, and 
$\tilde C(\bm H)$ is the binary linear code defined by $\bm H$, i.e., 
$
\tilde C(\bm H) := \{\bm{x}  \in \mathbb{F}_2^n \mid  \bm H \bm{x}^T = \bm{0}   \}.		
$
A binary to bipolar transform $b: \mathbb{F}_2 \rightarrow \{1, -1\}$
is defined as $b(0) := 1$ and $b(1) := -1$.
The bipolar code $C(\bm{H})$ is simply given by 
$$
 C(\bm H) := \{b(\bm{x}) \in \{1, -1\}^n \mid \bm{x}   \in \tilde C(\bm H) \}.
$$
The index sets $A(i)$ and $B(j)$ are defined as
$
	A(i) := \{j \mid j \in [n], H_{i, j} = 1 \} (i \in [m]) 
$
and 
$
	B(j) := \{i \mid i \in [m], H_{i, j} = 1   \} (j \in [n]),
$
respectively, where $H_{i,j}$ denotes the $(i,j)$-element of $\bm H$.
The notation $[n]$ represents the set $\{1,2, \ldots, n\}$.

\subsection{Definition of code-constraint polynomial}

The {\em code-constraint polynomial} for $C(\bm H)$ is a multivariate polynomial defined as
\begin{equation}
	h(\bm{x}) := \sum_{j = 1}^n (x_j^2 - 1)^2 
	+ \sum_{i = 1}^m \left( \left(\prod_{j \in A(i)} x_j \right)  - 1 \right)^2, 
\end{equation}
where $\bm x := (x_1, \ldots, x_n)\in \mathbb{R}^n$.
In this equation, the first term represents the bipolar constraint for $\bm{x} \in \{+1, -1\}^n$,
and the second term corresponds to the parity constraint induced by $\bm H$, i.e.,
if $\bm x \in C(\bm H)$, we have 
$$\left(\prod_{j \in A(i)} x_j \right) -1 = 0$$
 for any $i$.
Since the polynomial $h(\bm{x})$ has a {\em sum-of-squares} (SOS) form, the polynomial can be 
regarded as a penalty function that gives positive penalty values 
for non-codeword vectors in $\mathbb{R}^n$.
The code-constraint polynomial $h(\bm{x})$ is inspired by 
the non-convex parity constraint function used in the GDBF objective function \cite{wadayama10}.
The SOS form directly implies the most important property of $h(\bm{x})$, i.e., 
the inequality 
$
		h(\bm{x})  \ge 0
$
holds for any $\bm{x} \in \mathbb{R}^n$. The equality holds if and only if $\bm{x} \in C(\bm H)$.

\subsection{Gradient}
In the following discussion, we need the gradient of $h(\bm{x})$.
The first-order derivative of $h(\bm{x})$ with respect to $x_k (k \in [n])$ is given by
\begin{equation}
\frac{d}{d x_k}h(\bm{x}) 
		= 4 (x_k^2 - 1) x_k 
	+ \frac{2}{x_k} \sum_{i \in B(k) } \left(Q(i)^2 - Q(i) \right),
\end{equation}
where $Q(i)$ is defined as 
$Q(i) := \prod_{j \in A(i)} x_j.$
The gradient $\nabla h(\bm{x})$ is thus given by
\begin{equation}
	\nabla h(\bm{x}) = \left(\frac{d}{d x_1}h(\bm{x}), \ldots, \frac{d}{d x_n}h(\bm{x})     \right)^T.
\end{equation}

The point $\bm{x} \in \mathbb{R}^n$ satisfying the equality 
$
 \nabla h(\bm{x}) = \bm{0}
$
is a stationary point of $h$.
For any codeword $\bm{x} \in C(\bm H)$, 
$x_k^2 = 1$ for any $k \in [n]$ and 
$
	Q(i) = \prod_{j \in A(i)} x_j = 1
$
holds for any $i \in [m]$. This means $\nabla h = \bm{0}$.
Assume that a non-codeword bipolar vector 
$\bm{x} \in \{1, -1\}^n$ satisfying $\bm{x} \notin C(H)$ is given. 
For such $\bm{x}$, there exists a pair 
$(k, i)$ satisfying 
$
	 Q(i)^2  - Q(i) = 1 - (-1) = 2.
$
This implies that $\bm{x}$ is not a stationary point.
The above argument can be summarized as follows.
A codeword $\bm{x} \in C(\bm H)$ is a stationary point of $h$ and 
a non-codeword bipolar vector $\bm{x} \in \{1, -1\}^n, \bm{x} \notin C(\bm H)$
cannot be a stationary point.
A stationary point that is a codeword of $C(\bm H)$ is referred to as
a {\em codeword stationary point}.
Note that $h(\bm{x})$ can have non-codeword stationary points in general. 
For example, the zero vector $\bm 0 \in \mathbb{R}^n$ is an example of a non-codeword stationary point.

\subsection{Example of a code-constraint polynomial}

Suppose that the repetition code $$C := \{(+1, +1), (-1, -1) \}$$ is given.
The code-constraint polynomial for $C$ is thus given by
\begin{equation}
	h(\bm{x}) := (x_1^2 - 1)^2 + (x_2^2 - 1)^2 + (x_1 x_2 - 1)^2.
\end{equation}
The gradient of $h$ is given by 
	\begin{eqnarray}
		\nabla h(\bm{x}) =
	\left[
		\begin{array}{c}
		4 (x_1^2 - 1) x_1 + 2 (x_1 x_2 - 1) x_2 \\			
		4 (x_2^2 - 1) x_2 + 2 (x_1 x_2 - 1) x_1 \\			
		\end{array}
	\right].
	\end{eqnarray}
	
\begin{figure}[t]
\begin{center}
\includegraphics[width=\columnwidth]{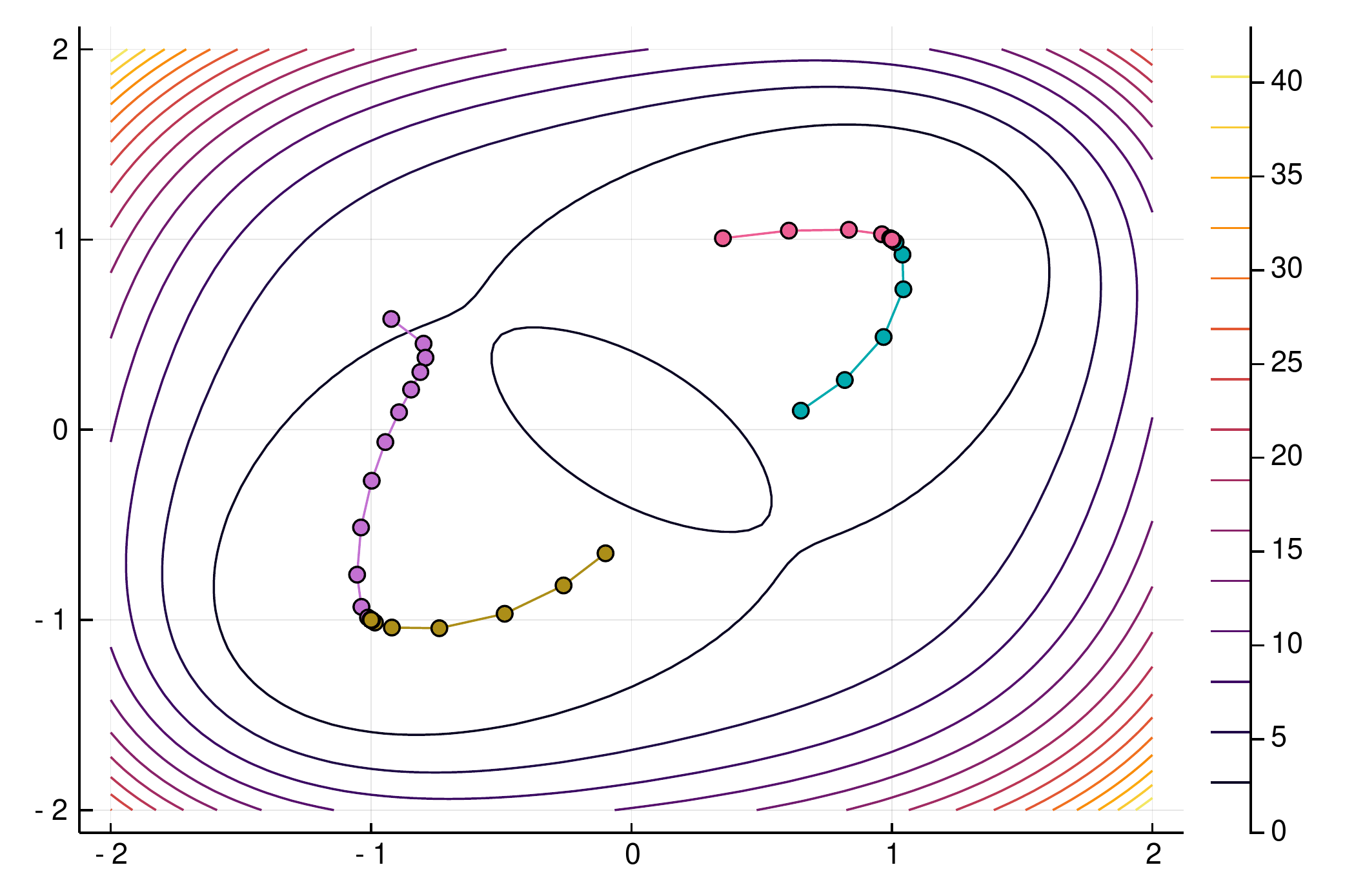}
\caption{Contour of the code-constraint polynomial $h(\bm{x}) = (x_1^2 - 1)^2 + (x_2^2 - 1)^2 + (x_1 x_2 - 1)^2$ for the repetition code of length. The trajectories of gradient descent processes (four different initial points) are depicted.}
\label{n2fig}
\end{center}
\end{figure}

In the case of repetition code of length 2,
there are three stationary points: 
$\{(1,1), (0,0), (-1, -1)\}$.
The two codeword points $\{(1,1), (-1, -1)\}$ are local minimums, and
$(0,0)$ is the local maximum.
If we apply a gradient descent method to $h(\bm{x})$, the gradient descent 
step is given by 
$
	\bm{x}^{(t+1)} = \bm{x}^{(t)} - \eta \nabla h(\bm{x}^{(t)}).	
$
Figure \ref{n2fig} illustrates several trajectories of gradient descent processes for 
different initial points.
We can observe that the trajectories generated by the gradient descent processes 
converge to the codewords $\{(+1, +1), (-1, -1) \}$.

As we can see from this example, the code-constraint polynomials as a multivariate 
function of $\bm x$ are non-convex and have several local minima and maxima in general. However, if the initial point is sufficiently close to a codeword stationary point, 
then the gradient descent step produces a convergent point sequence for the corresponding codeword. 
This {\em pull-in property} has critical importance in proximal decoding.

\section{Proximal decoding}

\subsection{Approximate maximum a posteriori decoding}

Assume that a sender transmits a codeword of $C(\bm H)$ to a given channel.
The channel is defined by a probability density function (PDF)
$p(\bm y |\bm x) (\bm x, \bm y \in \mathbb{R}^n)$.
The negative log-likelihood is defined as
$
L(\bm x;\bm y) :=	-\ln p(\bm{y}|\bm{x}).	
$
The MAP decoding rule is expressed as
\begin{equation}
\hat{\bm x} :=\text{argmin}_{\bm x \in C(\bm H)}\ 	L(\bm x;\bm y ) p(\bm{x}),
\end{equation}
where $p(\bm{x})$ is the prior PDF on the input space.
It is natural to make the equal probability assumption on $C(\bm H)$, which is given by 
\begin{equation}
p(\bm{x}) := \frac{1}{|C(\bm H)|}\sum_{\bm c \in C(\bm H)}\delta(\bm x - \bm c), 
\end{equation}
where $\delta$ is Dirac's delta function.
Instead of the true $p(\bm{x})$ above, here, we assume a prior PDF 
with the form 
\begin{align}
p(\bm{x}) = \frac{1}{Z} \exp\left(-\gamma h(\bm x) \right),	
\end{align}
where $Z$ is the normalizing constant and $\gamma$ is a positive constant.
Note that, at the limit $\gamma \rightarrow \infty$, we have
\begin{align}
({1}/{Z})\exp\left(-\gamma h(\bm x) \right) \rightarrow \frac{1}{|C(\bm H)|}\sum_{\bm c \in C(\bm H)}\delta(\bm x - \bm c).	
\end{align}
By substituting the result into $p(\bm x|\bm y)$, we immediately have 
\begin{equation}
p(\bm x| \bm y) \propto p(\bm y| \bm x) p(\bm x)	
 = \exp\left(- L(\bm x;\bm y) - \gamma h(\bm x) \right).
\end{equation}

The approximate MAP rule considered here is given by 
\begin{equation} \label{MAPrule}
\hat{\bm x} :=\text{argmin}_{\bm x \in \mathbb{R}^n}\ 	\left[ L(\bm x;\bm y )  +  \gamma h(\bm x) \right].
\end{equation}
The problem can be seen as a regression problem with a regularizer.
Note that the minimization problem (\ref{MAPrule}) is
also similar to the LASSO problem \cite{lasso} for sparse signal recovery.
The ISTA is derived from the LASSO formulation. It is natural to consider 
a counterpart of the ISTA for (\ref{MAPrule}), which is proximal decoding
to be presented in the next subsection.

\subsection{Proximal decoding}

Solving the approximate MAP problem (\ref{MAPrule})
can be seen as a non-convex regularized minimization problem.
In order to solve the approximate MAP problem efficiently, 
we will use the proximal gradient method \cite{proximal}.
The proximal operator of $f: \mathbb{R}^n \rightarrow \mathbb{R}$ is defined as
\begin{equation}
	\text{prox}_{f} (\bm{v})  := \text{argmin}_{\bm x \in \mathbb{R}^n} 
	\left(f(\bm{x}) + (1/2) \|\bm x - \bm v \|^2_2 \right),
\end{equation}
where $\|\cdot \|_2$ represents the Euclidean norm.
The proximal operators can be seen as a generalized projection.
The proximal operator $\text{prox}_{\gamma h} (\bm{x})$
can be well approximated (page 126 of \cite{proximal}) 
by a gradient descent step 
\begin{equation}
	\text{prox}_{\gamma h} (\bm{x}) \simeq \bm{x} - \gamma \nabla h(\bm{x}):= 
	{\cal P}_\gamma(\bm{x}),
\end{equation}
where the approximated proximal operator is said to be the {\em code-proximal operator}.

The proximal decoding proposed in the present paper is given by the following iterative 
process: 
\begin{eqnarray} \label{grad_step}
\bm{r}^{(k+1)} &:=& \bm{s}^{(k)} - \omega \nabla L(\bm s^{(k)};\bm y ) \\ \label{prox_step}
\bm{s}^{(k+1)} &:=& {\cal P}_\gamma(\bm{r}^{(k+1)}) = \bm{r}^{(k+1)} - \gamma \nabla h(\bm{r}^{(k+1)}),
\end{eqnarray}
for $k = 0, 1, 2, \ldots$, 
where $\omega$ is a positive number representing the step-size parameter of a 
gradient descent process in (\ref{grad_step}).
The step indicated by Eq. (\ref{grad_step}) is referred to as the gradient descent step, and the step indicated by Eq. (\ref{prox_step}) is said to be the code-proximal step.

\subsection{Box projection}
Let $B_{\eta}:= [-\eta, \eta]^n$, where $\eta$ is a positive constant 
slightly larger than one, be the $n$-dimensional hyper cube,
where $[a,b]:= \{x \in \mathbb{R} | a \le x \le b\}$.
The norm of the gradient $\|\nabla h(\bm{x})\|_2$ tends to be very large 
if $\bm x \notin B_{\eta}$ due to a property of the code-constraint polynomial. 
In the proximal decoding process defined above, this may cause numerical 
instability (oscillation or divergent behavior) in some cases.
In such a case, we can use 
\begin{equation}\label{box_proj}
\bm{s}^{(k+1)} := \Pi_\eta\left({\cal P}_\gamma(\bm{r}^{(k+1)})  \right).	
\end{equation}
instead of (\ref{prox_step}) in order to prevent numerical instability.
The projection operator $\Pi_\eta: \mathbb{R}^n \rightarrow \mathbb{R}^n$ represents the
projection onto $B_{\eta}$.

Let us discuss the time complexity per iteration of 
the proximal operation. For evaluating ${\cal P}_\gamma(\bm x)$, we require 
the gradient of $h(\bm{x})$. Let $k$ be the number of ones in $\bm H$.
All of the quantities $Q(i)$ for $i \in [m]$ can be calculated with time complexity $O(k)$. This means that the time complexity for evaluating the gradient of $h(\bm{x})$ is $O(n + k)$. If $C(\bm H)$ is an LDPC code, then $k = O(n)$ holds in general. This implies that the complexity for obtaining $\nabla h(\bm x)$ becomes $O(n)$, which is the practical time complexity, because $O(n)$ is the same as the complexity of belief propagation (BP) decoding for LDPC codes. 

\subsection{Proximal decoding for a massive MIMO channel}

The principle of the proximal decoding is applicable to any channel model 
if we precisely know the negative log-likelihood function $L(\bm x; \bm y)$  
and its gradient $\nabla L(\bm x; \bm y)$ can be efficiently evaluated.
In the present paper, we focus on a specific type of channel,
i.e., the LDPC-coded massive MIMO channel, which is of practical importance.
Let $\bm A \in \mathbb{R}^{m \times n}$ be a channel matrix. 
Suppose that a received word $\bm{y} \in \mathbb{R}^m$ is given by
\begin{align}
	\bm{y} = \bm{A} \bm{x} + \bm{w},	
\end{align}
where $\bm{w} \in \mathbb{R}^m$ is a Gaussian noise vector, the components of which follow an i.i.d. Gaussian distribution.
The channel input vector $\bm{x}$ is assumed to be a codeword of $C(\bm H)$,
which means that we assume BPSK modulation.
In this problem setting, the PDF representing the channel is given by
$$
p(\bm y |\bm x) = a \exp \left(- b \|\bm y - \bm A \bm x\|^2 \right),
$$
where $a$ and $b$ are positive constants.
We thus have the approximate MAP decoding problem for an LDPC-coded massive MIMO channel: 
\begin{equation}
\hat{\bm x} =
 \text{argmin}_{\bm x \in \mathbb{R}^n}\ \left[\|\bm y - \bm A \bm x\|^2 +  \gamma h(\bm x) \right].
\end{equation}
Since 
$
\nabla \|\bm y - \bm A \bm x\|^2  \propto \bm{A}^T (\bm A \bm x - \bm y)	,
$
an iteration of proximal decoding for LDPC-coded massive MIMO channels 
can be summarized as
\begin{eqnarray}
\bm{r}^{(k+1)} &=& \bm{s}^{(k)} - \omega \bm A^T (\bm A \bm{s}^{(k)} - \bm y)	 \\ 
\bm{s}^{(k+1)} &=& {\cal P}_\gamma(\bm{r}^{(k+1)}).
\end{eqnarray}
In the following experiments, we set $\bm{s}^{(k)} := \bm 0$.
However, there are alternative choices for the initial point, i.e., 
an estimate of the zero forcing detector or the MMSE detector can be 
used as an initial point.

\section{Numerical experiments}

\subsection{Proximal iteration for Hamming code}

\begin{figure}[t]
\begin{center}
\includegraphics[width=\columnwidth]{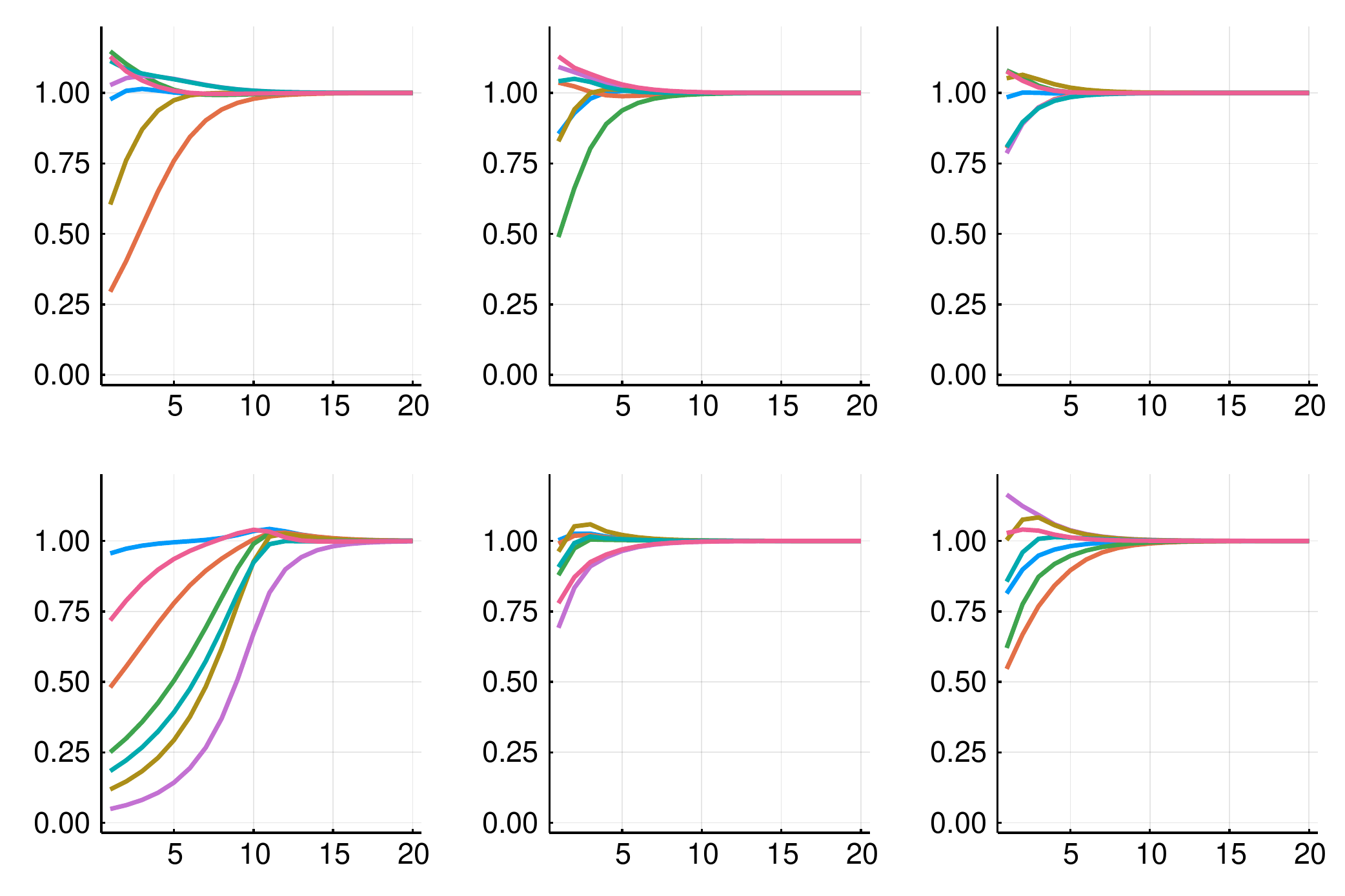}
\caption{Trajectories of the proximal iteration for the (7,4,3) Hamming code (six trials). The parameter $\gamma$ is set to $0.05$. Each curve in a graph corresponds to a component of $\bm x^{(k)}$. The initial value is set to $\bm y$.
}
\label{hamming}
\end{center}
\end{figure}

We start from an experiment to confirm the behavior of
a proximal iteration based on the code proximal operator ${\cal P}_\gamma$.
A simple proximal iteration \cite{proximal}
$
\bm{x}^{(k+1)} = {\cal P}_\gamma(\bm{x}^{(k)}), \ k = 0, 1, 2, \ldots 		
$
for $(7,4,3)$ Hamming code is examined.
The experimental setting is as follows.
We assume that the all-one codeword $\bm x := (1,1,1,1,1,1,1)$ 
is sent to an AWGN channel. The received word is given by
$\bm y = \bm x + \bm w$, where each i.i.d. component of $\bm w$
follows a Gaussian distribution with a mean of zero and a standard deviation of 0.5.
We used the received word $\bm y$ as the initial value of
the proximal iteration, i.e., $\bm x_0 := \bm y$.
The trajectories of components in ${\bm x}^{(k)}$ are depicted in 
Fig. \ref{hamming}. The horizontal axis represents the number of iteration 
steps, and the vertical axis indicates the value of $x_i^{(k)}, i \in [7]$. 
A curve in a graph corresponds to each component of $\bm x^{(k)}$.
From Fig. \ref{hamming}, we can observe that 
the proximal iteration promotes the convergence to 
the all-one codeword, i.e., the trajectories are attracted to the codeword point 
in these proximal iterations. The {\em pull-in property}
of the code proximal operator 
is confirmed based on this result.
Note that any codeword stationary point 
is a fixed point of ${\cal P}_\gamma$.

\subsection{LDPC-coded massive MIMO channel}

\subsubsection{Problem setup}
\label{prob_setup}
In this subsection, we follow the real-valued MIMO model 
discussed in \cite{tpg}. Let $\bm A' := \{a'_{i,j}\} \in \mathbb{C}^{M \times N}$ be 
a channel matrix, where $a_{i,j}'$ is the fading coefficient corresponding to the path between the $j$th transmit antenna and the $i$th receive antenna.
Here, we assume that each component of $\bm A'$  follows the {\em Kronecker model} \cite{Kronecker}, which is a simple channel model representing the spatial correlation between antenna elements.
Let $\rho (0 \le \rho < 1)$  be the spatial correlation factor. 
The correlation matrix for the receiver side is given by
$
\bm R_r := \{r_{i,j}\}_{1 \le i,j \le M}, 
r_{i,j}:=\rho^{|i - j|}
$
and the correlation matrix for a transmitter side is given by
$
\bm R_t := \{t_{i,j}\}_{1 \le i,j \le N}, 
t_{i,j}:=\rho^{|i - j|}. 
$
In the Kronecker model, a channel matrix $\bm A'$ is represented by 
$
\bm A' := \bm R_r^{1/2} \bm G (\bm R_t^{1/2} )^{T}, 
$
where each element of the matrix $\bm G \in \mathbb{C}^{M \times N}$
follows a complex circular Gaussian PDF with zero mean and unit variance.
Note that $\bm A'  = \bm G$ holds when there is no spatial correlation, i.e., $\rho = 0$.

We assume QPSK modulation for transmitted signals.
An equivalent real-valued MIMO model with BPSK modulation can be defined as
$\bm y = \bm A \bm x + \bm w$,
where $\bm A$ is given by
$$
	\bm A = 
	\left[
	\begin{array}{cc}
		Re(\bm A') & -Im(\bm A') \\
		Im(\bm A') & Re(\bm A') 
	\end{array}
	\right] \in \mathbb{R}^{m \times n}.
$$
Note that $m = 2M$ and $n = 2N$ holds. 
The transmitted word $\bm x$ is randomly 
chosen from $C(\bm H)$ according to the uniform distribution.
Each component of the noise vector 
$\bm w \in \mathbb{R}^m$ is an i.i.d. Gaussian PDF with zero mean and 
variance $\sigma_w^2/2$. In this model, $\sigma_w^2$ is related to 
the signal to noise ratio $\sf SNR$ by 
$
	\sigma_w^2 := (2N)/{\sf SNR}.
$
The details of the equivalence of the complex-valued model and 
the real-valued model can be found in \cite{tpg}.
In the following experiment, we used (3,6)-regular LDPC codes 
with $n =204$ and $m=102$.
The step-size parameter $\omega$ used in the gradient descent step 
is set to $\omega := 2.0/(\lambda_{min} + \lambda_{max})$, where 
$\lambda_{min}$ and $\lambda_{max}$ are the minimum and maximum 
eigenvalues of $\bm A^T \bm A$, respectively.
In the following experiments, we used 
the box projection (\ref{prox_step}) with $\eta = 1.5$ in the proximal step.

\subsubsection{Baseline schemes}

For the purpose of comparison, we exploited a proximal-based detection 
algorithm, referred to as a {\em Tanh detector}, given by the following recursion 
\cite{tpg}\cite{pic}:
\begin{align}
\bm{r}^{(k+1)} &= \bm{s}^{(k)} - \omega \bm  A^T (\bm A \bm x - \bm y),  \\
\bm{s}^{(k+1)} &= \tanh(\alpha \bm{r}^{(k+1)}),
\end{align}
where $\alpha$ is a positive real value. Furthermore, the MMSE detector defined as 
\begin{equation}\label{mmse}
\bm{\hat x} := \bm A^T (\bm A \bm A^T + (\sigma_w^2/2) \bm I)^{-1} \bm y	
\end{equation}
is also examined as a baseline scheme.

\subsubsection{Convergence behavior}
Let $\bm {\hat x}$ be the estimated word obtained 
from these detection algorithms. 
The performance measure used herein is the averaged error value 
$
\|\bm x - \text{\rm sign}(\bm{\hat x}) \|
$
where $\bm x$ is the transmitted word, and $\hat{\bm x}$ indicates an estimate 
obtained from the detector.

\begin{figure}[t]
\begin{center}
\includegraphics[width=\columnwidth]{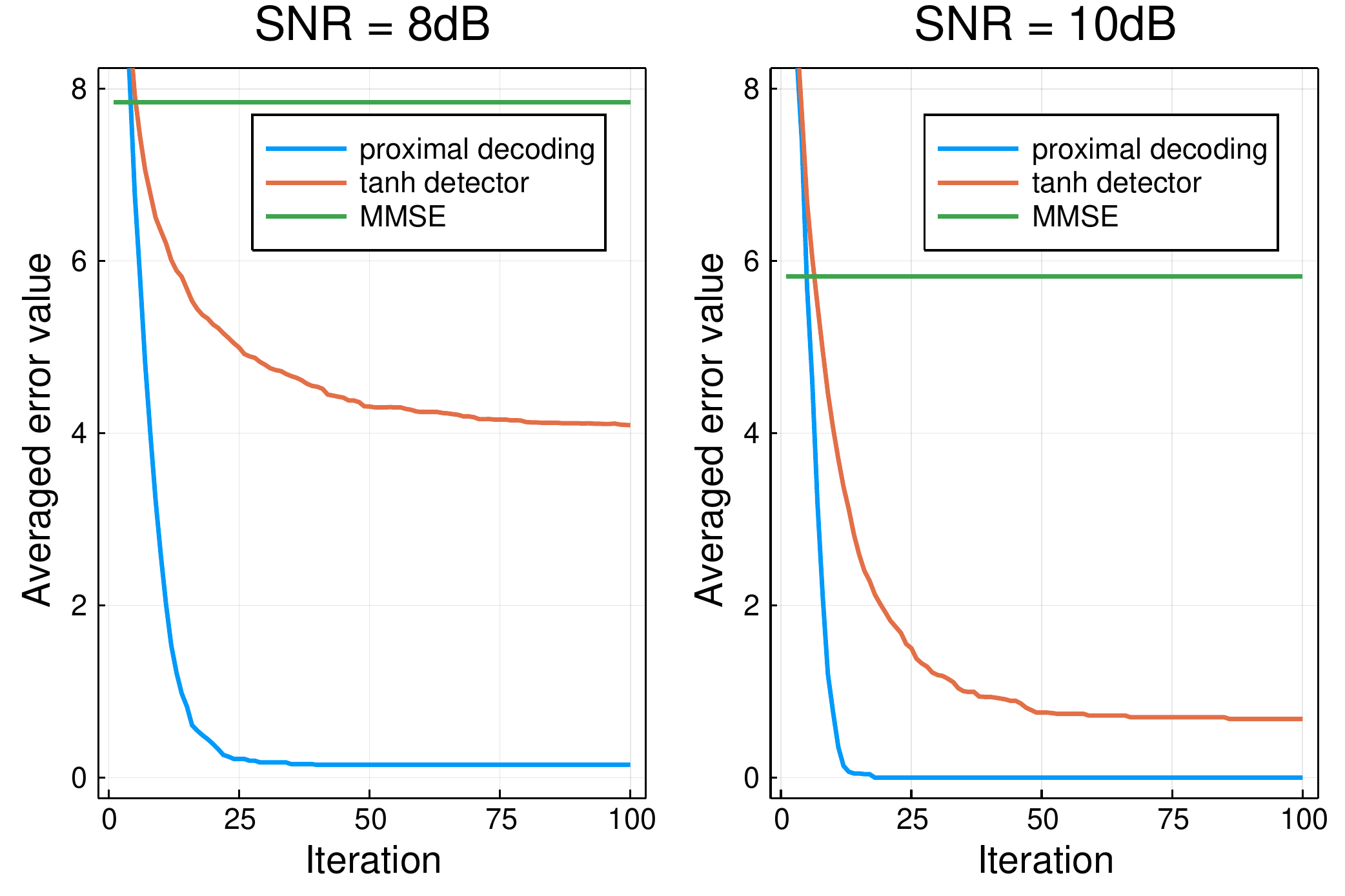}
\caption{Comparison of the averaged error value 
$\|\bm x - \text{\rm sign}(\bm{\hat x}) \|$ under $\rho = 0$ (no spatial correlation).
The number of received and transmitted antennas are $N = M = 102$.  (Left): {\sf SNR} = 8 (dB), (Right): {\sf SNR} = 10 (dB).
The error values are averaged for 100 trials. The choice of the parameter $\gamma$ is crucially important to obtain appropriate performance for proximal decoding. We chose $\gamma = 0.05$ for these experiments. 
	The parameter $\alpha$ used in the tanh detector is set to $2.0$. In all of the schemes, the step-size parameter is set to $\omega = 2.0/(\lambda_{min} + \lambda_{max})$.
}
\label{errorvals}
\end{center}
\end{figure}

Figure \ref{errorvals} presents the averaged error as a function of the number of iterations 
when there is no spatial correlation, i.e., $\rho = 0.0$.
Proximal decoding provides much smaller averaged error values and faster convergence compared with the Tanh detector and the MMSE detector.
Moreover, the saturated value of proximal decoding is much smaller than that of the Tanh detector.
These results imply that the parity constraint included in the code-constraint polynomial is fairly beneficial to obtain a reasonable solution.
We have also observed that the convergence speed of proximal decoding is sensitive to the choice of $\gamma$. In this experiment, $\gamma = 0.05$ provides the best result.
\subsubsection{Bit error rate performance}

The BER is the primal performance measure for the 
detection algorithms for massive MIMO systems. 
Here, we investigate the BER performance of 
proximal decoding and several benchmark schemes, such as the 
Tanh detector and the MMSE detector (with/without BP decoding).
The input of the BP decoder after the MMSE detection is 
set to $\xi \hat{\bm x}$, where $\xi$ is a positive constant,
and $\hat{\bm x}$ is an estimation vector obtained by (\ref{mmse}) 
without binary quantization.
The value of the scaling parameter $\xi$ is crucial for deriving the full performance 
of the BP decoding. In the following experiments, we set $\xi := 5$, which 
was heuristically adjusted. 
The channel model is the Kronecker model described in Subsection \ref{prob_setup}.

\begin{figure}[tbp]
\begin{center}
\includegraphics[width=\columnwidth]{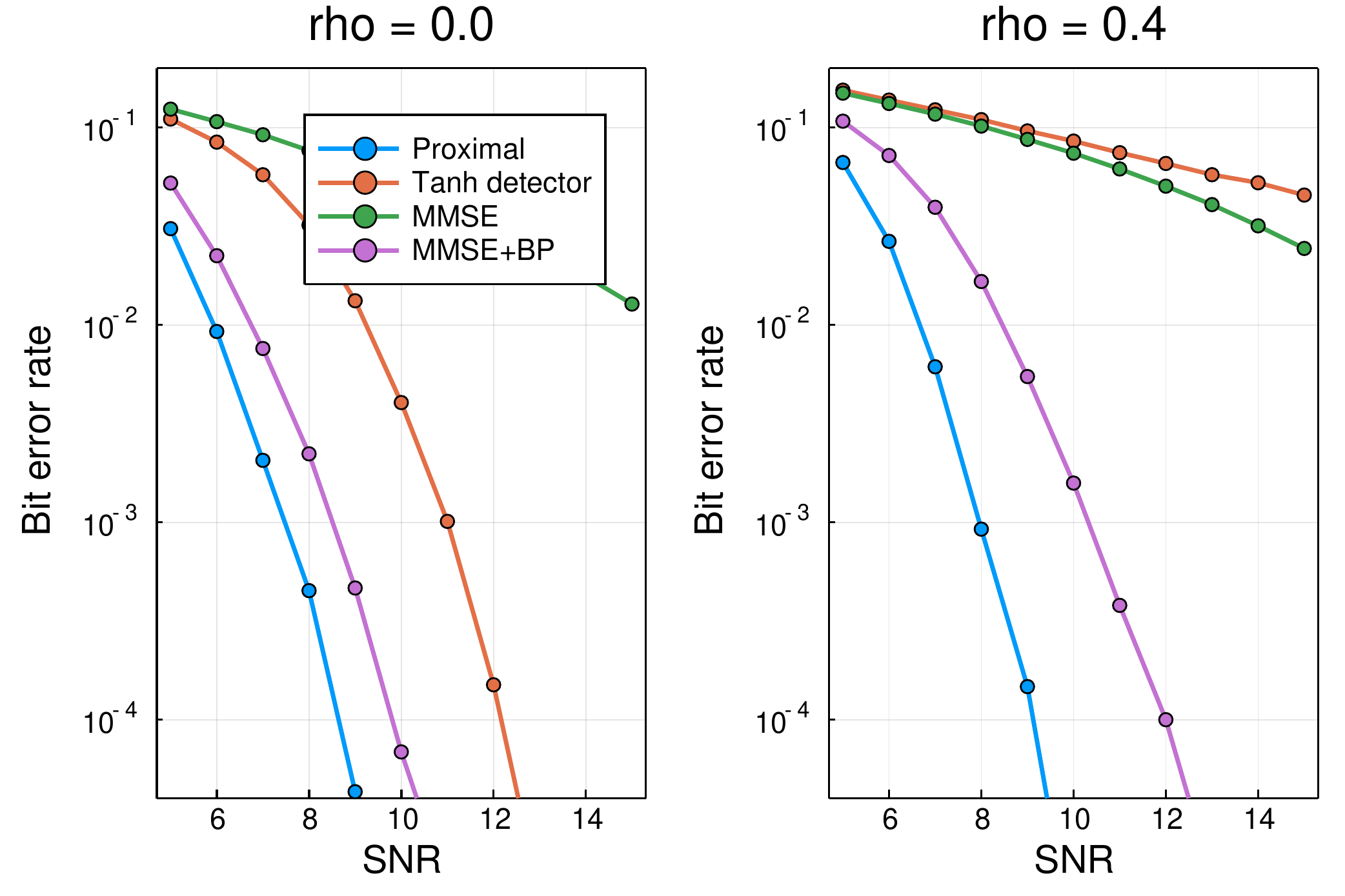}
\caption{Bit error rate performances of proximal decoding and baseline schemes. 
(Left) Without spatial correlation ($\rho = 0$); (Right) with spatial correlation ($\rho = 0.4$).
The number of received and transmitted antennas are $N = M = 102$. 
The error values are averaged for 5,000 trials. 
The parameter $\gamma$ used in the code proximal operator is set to $0.05$.
The step-size parameter is set to $\omega = 2.0/(\lambda_{min} + \lambda_{max})$.
The number of iterations for BP is 20, and the number of iterations for proximal decoding and the Tanh detector is set to 50. 
The scaling factor $\xi = 5$ is used for MMSE+BP.
}
\label{BER}
\end{center}
\end{figure}

Figure \ref{BER} presents the BER performances of the proposed and benchmark schemes.
The left-hand panel in Fig. \ref{BER} represents the case of no spatial correlation ($\rho = 0$), and the right-hand panel  indicates the results under the spatial correlation ($\rho = 0.4$).
Although the MMSE detector is the simplest detector among them, the error curve 
is not so steep in either Fig. \ref{BER} (Left) or Fig. \ref{BER} (Right).  Furthermore, the MMSE 
detector involves the inversion of a matrix requiring time complexity $O(n^3)$,
which is not negligible in terms of complexity in a massive MIMO scenario.
The combination of the MMSE detector followed 
by a BP decoder (MMSE+BP) is a standard and practical configuration 
of a receiver for LDPC-coded massive MIMO channels. 

The BER performance of
MMSE + BP provides a much steeper error curve as compared with the 
plain MMSE error curve. The Tanh detector also achieves much smaller BERs as compared with the naive MMSE detector when $\rho = 0$.
Compared with the Tanh detector and the MMSE (with/without BP decoding),
the BERs of proximal decoding
yield the smallest BERs.
In particular, the margin between the proposed method and MMSE + BP  is 
approximately 3 dB at BER $= 10^{-4}$ in Fig. \ref{BER}(Right).
Comparing Figs. \ref{BER}(Left) and \ref{BER}(Right), the performance of MMSE+BP deteriorates as $\rho$ increases.
The proposed method provides similar BER performances in both cases.

Although a number of studies have discussed joint detection and decoding for LDPC-coded MIMO channels, such as \cite{Hwang}, their time and circuit complexities are much larger than those for proximal decoding.
The complexity of proximal decoding is $O(\ell n^2)$, where $\ell$ represents the number of iterations, which is lower than the complexity of the MMSE detector if $\ell$ is constant.
Due to the pull-in property 
of the code proximal operator, a search point may be attracted by a codeword stationary point 
in a decoding process. This attractive force would achieve superior performance of the proposed 
method. 


\section{Summary}

In the present paper, we present proximal decoding as an instance of the 
approximate MAP decoding for LDPC codes.
Through numerical experiments, proximal decoding is shown to be competitive with 
known detection methods, such as MMSE + BP.
Although we restricted our attention to the case of LDPC-coded massive MIMO channels, the concept of proximal decoding can be applied to another non-trivial channel if the gradient of the negative log-likelihood function can be efficiently evaluated.
The approach presented in the present paper may open a new direction of optimization-based decoding algorithms.
Another preferable feature of proximal decoding is that all the subprocesses of proximal decoding are differentiable. 
Thus, we can apply standard deep learning techniques to optimize the internal parameters 
for achieving better performance.
Such a methodology, often referred to as {\em deep unfolding}
\cite{LISTA} \cite{TISTA}  \cite{Com2} \cite{TW19}, 
appears promising for tuning parameters $\gamma$ and $\omega$ in proximal decoding, 
which are highly influential in the case of the BER performance.

\section*{Acknowledgement}
The present study was supported  by a JSPS Grant-in-Aid for Scientific Research (B) Grant Number 19H02138 (TW).

\end{document}